\documentclass[conference,letterpaper]{IEEEtran}
\IEEEoverridecommandlockouts

\usepackage{cite}
\usepackage{amsmath,amssymb,amsfonts}
\usepackage{amsthm}
\usepackage{mathtools}
\usepackage{mathpartir}
\usepackage{stmaryrd}
\usepackage{algorithm}
\usepackage{algpseudocode}

\theoremstyle{definition}

\theoremstyle{plain}

\newtheorem{property}{Property}[section]

\theoremstyle{remark}

\newcommand{\anglebrackets}[1]{{\langle}#1{\rangle}}

\usepackage{graphicx}
\usepackage{textcomp}
\usepackage{xcolor}
\def\BibTeX{{\rm B\kern-.05em{\sc i\kern-.025em b}\kern-.08em
    T\kern-.1667em\lower.7ex\hbox{E}\kern-.125emX}}

\makeatletter
\newcommand{\linebreakand}{%
  \end{@IEEEauthorhalign}
  \hfill\mbox{}\par
  \mbox{}\hfill\begin{@IEEEauthorhalign}
}
\makeatother

\usepackage[top=0.75in, bottom=1.07in, left=0.625in, right=0.625in]{geometry}
\begin{document}

\title{Decentralized Multi-Agent System with \\Trust-Aware Communication
\thanks{This is the preprint version of the conference paper "Decentralized Multi-Agent System with Trust-Aware Communication" (Best Paper Award) in \textit{2025 IEEE International Symposium on Parallel and Distributed Processing with Applications (ISPA)}.}
}

\author{\IEEEauthorblockN{1\textsuperscript{st} Yepeng Ding}
\IEEEauthorblockA{
\textit{Hiroshima University}\\
Higashihiroshima, Japan \\
yepengd@acm.org}
\and
\IEEEauthorblockN{2\textsuperscript{nd} Ahmed Twabi}
\IEEEauthorblockA{
\textit{Hiroshima University}\\
Higashihiroshima, Japan \\
iman-twabi@hiroshima-u.ac.jp}
\and
\IEEEauthorblockN{3\textsuperscript{rd} Junwei Yu}
\IEEEauthorblockA{
\textit{The University of Tokyo}\\
Tokyo, Japan \\
yujw@satolab.itc.u-tokyo.ac.jp}
\linebreakand
\IEEEauthorblockN{4\textsuperscript{th} Lingfeng Zhang}
\IEEEauthorblockA{
\textit{The University of Tokyo}\\
Tokyo, Japan \\
zhang-lingfeng936@g.ecc.u-tokyo.ac.jp}
\and
\IEEEauthorblockN{5\textsuperscript{th} Tohru Kondo}
\IEEEauthorblockA{
\textit{Hiroshima University}\\
Higashihiroshima, Japan \\
tkondo@hiroshima-u.ac.jp}
\and
\IEEEauthorblockN{6\textsuperscript{th} Hiroyuki Sato}
\IEEEauthorblockA{
\textit{National Institute of Informatics}\\
Tokyo, Japan \\
schuko@nii.ac.jp}
}

\maketitle

\begin{abstract}
The emergence of Large Language Models (LLMs) is rapidly accelerating the development of autonomous multi-agent systems (MAS), paving the way for the Internet of Agents. However, traditional centralized MAS architectures present significant challenges, including single points of failure, vulnerability to censorship, inherent scalability limitations, and critical trust issues. We propose a novel Decentralized Multi-Agent System (DMAS) architecture designed to overcome these fundamental problems by enabling trust-aware, scalable, and censorship-resistant interactions among autonomous agents. Our DMAS features a decentralized agent runtime underpinned by a blockchain-based architecture. We formalize a trust-aware communication protocol that leverages cryptographic primitives and on-chain operations to provide security properties: verifiable interaction cycles, communication integrity, authenticity, non-repudiation, and conditional confidentiality, which we further substantiate through a comprehensive security analysis. Our performance analysis validates the DMAS as a scalable and efficient solution for building trustworthy multi-agent systems.
\end{abstract}

\begin{IEEEkeywords}
multi-agent system, blockchain, agent-to-agent, large language model, self-sovereign identity.
\end{IEEEkeywords}

\section{Introduction}
\label{sec:introduction}

The rapid advancements in Large Language Models (LLMs) \cite{zhao_survey_2023,yao_react_2023,touvron_llama_2023,chang_survey_2024} have opened unprecedented avenues for creating highly autonomous and intelligent agents. These LLM-augmented agents possess remarkable capabilities in understanding natural language, performing complex reasoning, planning intricate sequences of actions, and engaging in sophisticated communication. As the complexity of real-world problems outgrows the capacity of single agents, the paradigm is naturally shifting towards Multi-Agent Systems (MAS) \cite{van_der_hoek_multi-agent_2008,han_llm_2024,guo_large_2024}, where multiple agents collaborate to achieve shared or individual goals through interaction and coordination. This evolution envisions the Internet of Agents (IoA) \cite{chen_internet_2024}, a future where a vast network of interoperable AI agents can discover, interact with, and provide services to each other and human users across diverse domains.

While the promise of such interconnected agent ecosystems is immense, current practical implementations of LLM-augmented MAS often suffer from fundamental limitations inherent to centralized architectures \cite{yu_textual_2024,das_security_2025}. Relying on central servers or authorities for agent discovery, communication, and trust management introduces several critical vulnerabilities: a \emph{single point of failure} leading to system fragility, susceptibility to \emph{censorship and arbitrary control} by a central entity, significant \emph{scalability bottlenecks} as the number of agents and complexity of interactions grow, and pervasive \emph{trust issues}. In a centralized MAS, users and agents must implicitly trust the central operator to ensure data integrity, privacy, fair resource allocation, and honest service provision. Furthermore, verifying compliance with agreed-upon protocols or holding agents accountable for their outputs becomes challenging without transparent and immutable records. These challenges collectively hinder the realization of a robust, resilient, and equitable IoA.

Motivated by addressing these challenges, we introduce a novel \textbf{Decentralized Multi-Agent System (DMAS)} architecture. Our proposed DMAS leverages blockchain primitives \cite{wood_ethereum_2014,ding_bloccess_2023,garay_bitcoin_2024} and the principles of self-sovereign identity \cite{ding_leveraging_2022,ding_self-sovereign_2022,ding_data_2024} to enable trust-aware, scalable, and censorship-resistant interactions among autonomous agents. By decentralizing core functionalities, the DMAS provides a foundational framework for agents to operate and collaborate without reliance on a single, trusted third party. Our main contributions are summarized as follows:
\begin{itemize}
    \item We propose a decentralized multi-agent system architecture that integrates a \emph{Decentralized Agent Runtime} and a \emph{Trust-Aware Communication Protocol} to secure agent discovery and interactions while preserving performance.
    \item We provide a semi-formal analysis of DMAS’s security properties.
    \item We present a comprehensive performance analysis of the DMAS, demonstrating its high scalability and comparable efficiency with the centralized multi-agent system.
\end{itemize}

\section{Related Work}

The integration of large language models (LLMs) into multi-agent systems (MAS) has substantially advanced the sophistication of inter-agent collaboration. Recent surveys emphasize that LLM-augmented MAS harness the collective intelligence of multiple specialized agents to achieve capabilities that surpass those of individual agents, effectively addressing challenges such as hallucinations and single points of failure \cite{guo_large_2024}. By engaging in dialogue, debate, and mutual verification, multiple LLM agents emulate the cooperative dynamics of human teams in complex problem-solving, leading to improved reasoning accuracy and task performance \cite{wu_autogen_2024}.

Recent research has increasingly focused on reflective coordination paradigms, wherein agents collaboratively regulate their behavior through self-evaluation and continual learning \cite{huang_large_2023}. For instance, COPPER \cite{bo_reflective_2024} introduces a self-reflection mechanism, either embedded within individual agents or implemented via a shared mediator—that iteratively refines agent prompts using a learned reflector model. This model is trained via counterfactual reinforcement learning (specifically, Proximal Policy Optimization), which rewards agents for generating useful reflections. As a result, the system dynamically optimizes role assignments and interaction patterns, mitigating credit-assignment issues by attributing success to effective reflection rather than static agent roles.

LLM-based multi-agent systems have demonstrated effectiveness across diverse domains, including software engineering \cite{nam_using_2024,wu_survey_2024} and medical science \cite{thirunavukarasu_large_2023,singhal_large_2023}, showcasing their potential for tackling complex, interdisciplinary tasks that require long-term planning and specialized knowledge \cite{guo_large_2024,hong_metagpt_2024}. However, many of these applications adopt centralized coordination architectures, which introduce security vulnerabilities and limit scalability, particularly in untrusted or decentralized environments.

\section{Decentralized Multi-Agent System}

We formulate a decentralized multi-agent system (DMAS) to enable trust-aware and scalable interactions among autonomous agents without reliance on centralized control.

\subsection{Decentralized Agent Runtime}

\begin{figure}[htbp]
\centerline{\includegraphics[width=0.65\columnwidth]{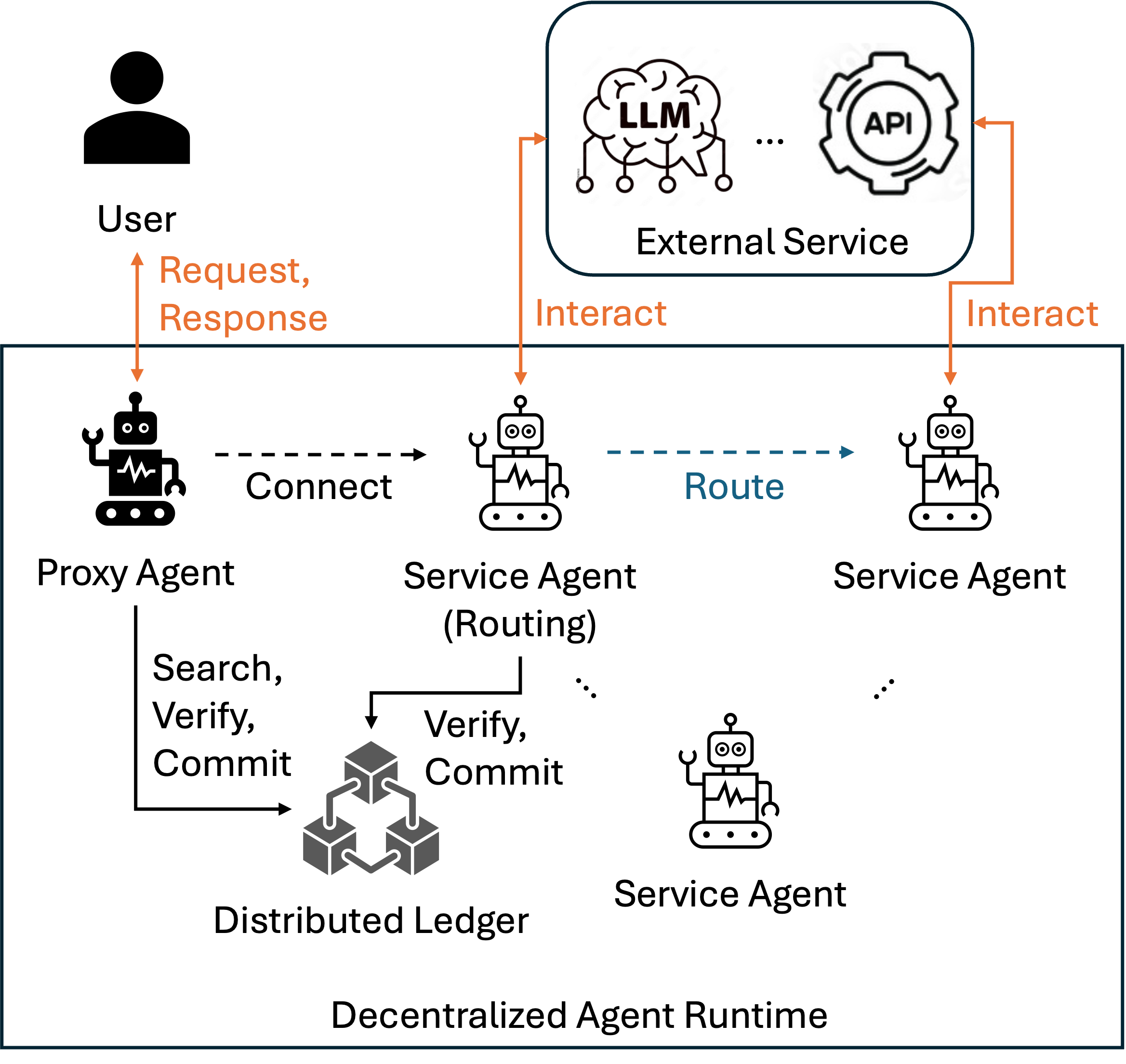}}
\caption{Overview of the decentralized agent runtime.}
\label{fig:overview}
\end{figure}

The DMAS introduces a decentralized runtime architecture to facilitate secure and scalable agent-to-agent interactions. This architecture, as depicted in Figure~\ref{fig:overview}, comprises proxy agents and service agents, disseminated across decentralized environments.

\subsubsection{Agent Type}

The runtime primarily comprises two agent types: Proxy Agents (PAs) and Service Agents (SAs). These agents operate across both on-chain and off-chain environments, establishing a robust and privacy-preserving communication paradigm.

\paragraph{Proxy Agents}
PAs serve as the user's primary interface to the decentralized agent runtime, acting as intelligent intermediaries responsible for routing user requests to appropriate SAs based on their advertised capabilities. Each user within the DMAS ecosystem is mandated to host a dedicated Proxy Agent. PAs are designed to handle complex user queries by potentially orchestrating interactions with multiple SAs.

\paragraph{Service Agents}
SAs form the computational backbone of the DMAS. They are designed to execute specific tasks or provide specialized services. SAs are responsible for processing requests originating from either PAs or other SAs and generating responses commensurate with their defined capabilities. Critically, SAs operate autonomously off-chain, performing computations or accessing external data sources as required by their service definitions. This might include accessing real-world APIs, performing complex data analysis, or executing machine learning models.

\subsubsection{Verifiable Agent Registry}

We introduce Self-Sovereign Identity (SSI) principles for robust agent identity and capability management within the DMAS. Correspondingly, we construct a Verifiable Agent Registry (VAR) implemented as a smart contract operating on blockchains. The VAR is central to establishing trust and transparency in agent interactions by providing a public, immutable record of agent identities and capabilities.

\paragraph{DID Management}

Each PA and SA is assigned a unique Decentralized Identifier (DID). These DIDs are cryptographically verifiable and managed by a Verifiable Data Registry (VDR). The VDR is implemented as a smart contract on a public blockchain (e.g., Ethereum, Polygon), ensuring immutability, transparency, and censorship resistance in DID registration, updates, and revocation. The VDR acts as a public, tamper-proof directory for all registered SAs, providing a foundational layer of trust, which aligns with blockchain's core tenets of decentralization, guaranteeing that an agent's identity cannot be arbitrarily altered or censored.

\paragraph{DID Resolution}

A core function of the VDR is to enable the resolution of an SA's DID to a corresponding capability schema. This schema, structured in JSON-LD, contains a machine-readable description outlining the services, functionalities, input/output parameters, and operational parameters offered by the SA. This on-chain resolution mechanism provides a trusted and verifiable source of information regarding agent capabilities, which is crucial for informed peer discovery and secure interaction initiation. The cryptographic link between the DID and its associated capability schema ensures that the advertised capabilities are genuinely linked to the registered agent, preventing impersonation or misrepresentation and fostering a secure decentralized runtime.

\subsubsection{Agent Hosting}

While DIDs and their resolution occur on-chain, the SAs themselves, along with their computational capabilities and execution environments, are hosted off-chain. This design choice optimizes for scalability and performance, preventing the blockchain from becoming a bottleneck for heavy computational tasks and enabling complex, long-running services. An SA, once its DID is resolved to its capability schema, can be effectively routed by a PA using the connection information (e.g., IP address, API endpoint, communication protocol details) embedded within that schema. This off-chain communication leverages established network protocols (e.g., HTTP/S, WebSockets) for efficient data exchange, highlighting the hybrid nature of the DMAS. The blockchain acts as a trust anchor and coordination layer, while the actual heavy lifting is performed off-chain.

\subsubsection{Agent Memory}
Due to the inherent space sensitivity and cost associated with on-chain storage, memory management within the DMAS is exclusively enforced on the PAs. PAs are responsible for maintaining conversational context and assembling received responses from SAs off-chain. This accumulated context is then utilized to inform and contextualize subsequent requests to SAs, enabling multi-turn interactions and complex task orchestration, similar to how an LLM maintains conversational state. Conversely, SAs are designed to be memoryless (stateless) from the perspective of persistent conversational history. Each request processed by an SA is treated as an independent operation, with all necessary context provided by the requesting PA. Our stateless design for SAs simplifies their implementation, enables parallel processing, and minimizes their on-chain footprint, which contributes to the overall scalability and efficiency of the DMAS.

\subsection{Trust-Aware Protocol}

We formulate a trust-aware protocol for the DMAS to delineate the formalized interaction mechanisms between a user's designated Proxy Agent, denoted as $u$, and a collective of autonomous Service Agents, represented as the set $S = \{s_1, s_2, \dots, s_m\}$. This protocol is meticulously designed to imbue the DMAS with properties of verifiable trust, accountability, and conditional data exchange, leveraging the immutability and transparency afforded by a distributed ledger, $\mathfrak{B}$. The entire interaction sequence is predicated upon cryptographic primitives and on-chain attestations, thereby minimizing reliance on centralized authorities and fostering true decentralization.

\subsubsection{Service Discovery}

A service discovery process is initiated by a user request, denoted by $\triangleright$, which is handled by a PA $u$. Agent $u$ is tasked with continuously communicating with a set of suitable SAs capable of fulfilling the requirements specified by $\triangleright$, until a termination condition $\tau$ is satisfied. This process goes beyond simple lookup, involving dynamic interaction and delegation.

To address the scalability challenges posed by the potentially large cardinality of $S$, the service discovery mechanism employs a delegation pattern. In this pattern, each SA is equipped with the ability to forward incoming requests to other candidate SAs, based on the user request $\triangleright$ and the contextual state $\Gamma(u)$ maintained by the PA $u$. If an SA responds with a directive to forward the request, we classify the result as non-terminal; otherwise, if the response terminates further routing, it is classified as terminal. This delegation allows for a flexible and dynamic discovery process, where specialized SAs can guide the PA to more relevant services. We present two common discovery strategies based on the delegation pattern: depth-first strategy and breadth-first strategy, though the discovery strategy can be customized by PAs based on specific application requirements or heuristic rules.

\paragraph{Depth-First Strategy}
The depth-first strategy is formalized in Algorithm~\ref{alg:depth_first}. Initially, $u$ selects a small candidate set of SAs via $\textit{FirstSelect}(u, \triangleright)$, a predefined method such as a hardcoded list, semantic similarity techniques (e.g., LLM-based matching for richer understanding of service descriptions), or reputation scores. It then engages in iterative communication with each selected SA, recursively following routing suggestions until either no further candidates remain or the termination condition $\tau$ is satisfied. The communication procedure between $u$ and an SA $s$, denoted $\textit{Com}(u, s)$, is described in detail in Section~\ref{sec:verifiable_com}. This strategy is suitable for scenarios where a deep, specialized exploration of a service chain is preferred.

\begin{algorithm}
\caption{Depth-First Discovery Strategy}\label{alg:depth_first}
\begin{algorithmic}
\Require $u$, $\triangleright$, $\tau$
\Ensure $\vec{R}$: an ordered list of terminal responses
\State $\vec{R} \gets \emptyset$
\State $\hat{S} \gets \textit{FirstSelect}(u, \triangleright)$ \Comment{LIFO stack with initial candidates}

\While{$|\hat{S}| > 0 \land \tau \implies \bot$}
    \State $s \gets \textit{Pop}(\hat{S})$
    \State $r \gets \textit{Com}(u, s)$ \Comment{$u$ communicates with $s$}
    \If{$\textit{IsTerminal}(r) = \top$}
        \State $R \gets R \cup \{ r \}$
    \Else
        \State $\textit{Push}(\hat{S}, \textit{SA}(r))$ \Comment{Add forwarded candidates to stack}
    \EndIf
\EndWhile

\State \Return $\vec{R}$
\end{algorithmic}
\end{algorithm}

\paragraph{Breadth-First Strategy}
In contrast to the depth-first strategy, the breadth-first discovery strategy explores the SA space in a level-wise manner via a queue. The PA $u$ initiates the discovery process with an initial set of candidate SAs, selected via $\textit{FirstSelect}(u, \triangleright)$. Rather than following each routing path deeply, this strategy systematically explores all SAs at the current level before proceeding to the next. The process proceeds iteratively until the termination condition $\tau$ is met or no further candidates remain in the queue. This approach can be beneficial for discovering a wider range of potentially relevant services quickly, suitable for scenarios requiring diverse responses or exploring many options concurrently.

\paragraph{Termination Condition}
The termination condition $\tau$ governs when the service discovery process should halt. Rather than being a fixed Boolean value, $\tau$ is formulated as a predicate function maintained by the PA $u$. We define $\tau$ as a conjunction or disjunction over a configurable set of atomic predicates, such as: fulfillment of the request (e.g., "received at least three viable options"), the number of SA communications exceeding a predefined threshold, expiration of a wall-clock timeout, or an explicit human intervention signal.

\subsubsection{Verifiable Communication}
\label{sec:verifiable_com}

The DMAS leverages a distributed ledger, $\mathfrak{B}$, to enable verifiable communications between PAs and SAs. The communication follows a hybrid protocol that records requests and response receipts on-chain, ensuring verifiability and preventing malicious attacks such as repudiation or tampering, which is critical for building trust in a decentralized environment where agents may not inherently trust each other. This protocol comprises three core steps: request commitment, response commitment, and response retrieval, involving the resolution of DIDs via the VAR residing on $\mathfrak{B}$ to yield capability schemas for available SAs. 

Let $\textit{DID}(s)$ denote the DID of SA $s \in S$, and $\textit{CAP}(\textit{DID}(s))$ denote its resolved capability schema. We assume a communication initiated by a PA $u$ to an SA $s \in S$.

\paragraph{Request Commitment}

The PA $u$ constructs an off-chain request payload, $\mathcal{P}(\triangleright)$, intended for $s$. To establish an immutable record of the request initiation and to ensure its integrity, $u$ simultaneously broadcasts a request transaction, $\mathcal{X}(\mathcal{P}(\triangleright)) \triangleq \anglebrackets{\textit{DID}(u), \textit{DID}(s), \mathcal{H}(\mathcal{P}(\triangleright))}$, to the distributed ledger $\mathfrak{B}$. $\mathcal{H}(\mathcal{P}(\triangleright))$ is a cryptographic hash of the off-chain request payload, acting as a unique fingerprint of the request.

Upon the successful inclusion of $\mathcal{X}(\mathcal{P}(\triangleright))$ into $\mathfrak{B}$, $u$ transmits the detailed off-chain payload $\mathcal{P}(\triangleright)$ and a reference to $\mathcal{X}(\mathcal{P}(\triangleright))$ (e.g., the transaction hash or block number) to $s$ via a secure, direct communication channel (e.g., end-to-end encrypted peer-to-peer connection). This ensures that the SA receives both the content of the request and verifiable proof of its on-chain commitment, preventing $u$ from later denying having sent the request.

\paragraph{Response Commitment}

Upon receipt of $\mathcal{P}(\triangleright)$, SA $s$ first verifies its authenticity and validity by querying $\mathfrak{B}$ to confirm the existence and integrity of $\triangleright$ corresponding to $\mathcal{H}(\mathcal{P}(\triangleright))$ in the on-chain request commitment. If $\mathcal{H}(\mathcal{P}(\triangleright))$ matches, $s$ proceeds with the execution of the requested service. Let $\triangleleft$ denote the raw response generated by $s$.

To ensure conditional access and privacy-preserving delivery of $\triangleleft$, $s$ performs the following operations:
\begin{enumerate}
    \item Encrypt the raw response using a symmetric encryption key, $\kappa$: $\bar{\triangleleft} = \textit{Enc}(\triangleleft, \kappa)$.
    \item Store $\bar{\triangleleft}$ in an off-chain storage facility, denoted as $\mathcal{D}(\bar{\triangleleft})$.
    \item Construct a response transaction, $\mathcal{X}(\triangleleft)$, and broadcasts it to $\mathfrak{B}$, where
    $\mathcal{X}(\triangleleft) \triangleq \anglebrackets{\mathcal{H}(\mathcal{X}(\mathcal{P}(\triangleright))), H(\bar{\triangleleft}), \eta}$, and $\eta$ specifies the conditions under which $u$ may obtain the decoding key $\kappa$ (e.g., payment of a service fee).
    \item Notify $u$ with the storage location $\mathcal{D}(\bar{\triangleleft})$ and the transaction $\mathcal{X}(\triangleleft)$ via a secure, direct communication channel.
\end{enumerate}

\paragraph{Response Retrieval}
Upon receiving the notification for $\mathcal{X}(\triangleleft)$, the PA $u$ retrieves the encrypted response $\bar{\triangleleft}$ via $\mathcal{D}(\bar{\triangleleft})$ and follows the operations below:
\begin{enumerate}
    \item $u$ verifies the integrity of $\bar{\triangleleft}$ by comparing its cryptographic hash with $H(\bar{\triangleleft})$ recorded on $\mathfrak{B}$. This ensures that the retrieved encrypted response has not been tampered with during transit or storage;
    \item $u$ proceeds to satisfy $\eta$. For instance, if $\eta$ involves a payment, $u$ initiates a payment transaction on $\mathfrak{B}$ to the specified wallet address in $\eta$. The smart contract managing $\eta$ verifies the fulfillment of the conditions;
    \item Once $\eta$ are verifiably fulfilled on $\mathfrak{B}$ (e.g., confirmed by the smart contract or by a verifiable credential), $u$ notifies $s$ via a secure, direct communication channel to retrieve $\bar{\kappa} = \textit{Enc}(\kappa, \textit{PK}(u))$, where $\textit{PK}(u)$ is the public key of $u$. This ensures that only the intended PA, whose public key is known, can decrypt the symmetric key, maintaining confidentiality.
    \item Finally, $u$ decrypts $\bar{\kappa}$ with its private key $\textit{SK}(u)$ to obtain $\kappa$, and then reveals $\triangleleft$ (the raw response) by decrypting $\bar{\triangleleft}$ using $\kappa$:
    $$\triangleleft = \textit{Dec}(\bar{\triangleleft}, \textit{Dec}(\bar{\kappa}, \textit{SK}(u)))$$
\end{enumerate}

This multi-step response retrieval process ensures verifiability, trust, and privacy throughout the interaction lifecycle, providing a robust framework for secure and auditable agent communications within a decentralized environment. It mitigates common risks associated with off-chain interactions by anchoring critical steps to the immutable ledger.

\section{Security Analysis}
\label{sec:security_analysis}

We provide a semi-formal security analysis of the DMAS regarding its key security properties. 

\begin{property}[Communication Integrity]
    The content of requests $\mathcal{P}(\triangleright)$ and responses $\triangleleft$ remains unaltered during transmission and storage between a PA $u$ and a SA $s$.
\end{property}

\begin{proof}
We prove the communication integrity regarding the request and response integrity, respectively.
For \textbf{Request Integrity},
    \begin{enumerate}
        \item When $u$ initiates a request, it computes a cryptographic hash $\mathcal{H}(\mathcal{P}(\triangleright))$ of the off-chain request payload $\mathcal{P}(\triangleright)$.
        \item $\mathcal{H}(\mathcal{P}(\triangleright))$ is then committed on $\mathfrak{B}$ as part of the request transaction $\mathcal{X}(\mathcal{P}(\triangleright)) = \anglebrackets{\textit{DID}(u), \textit{DID}(s), \mathcal{H}(\mathcal{P}(\triangleright))}$. The immutability of $\mathfrak{B}$ ensures that this committed hash cannot be altered post-recording.
        \item Upon receiving $\mathcal{P}(\triangleright)$ off-chain, $s$ independently computes its hash, $\mathcal{H}'(\mathcal{P}(\triangleright))$, and queries $\mathfrak{B}$ to retrieve the original committed hash $\mathcal{H}(\mathcal{P}(\triangleright))$.
        \item $s$ verifies the integrity of the received payload by comparing $\mathcal{H}'(\mathcal{P}(\triangleright))$ with $\mathcal{H}(\mathcal{P}(\triangleright))$. If $\mathcal{H}'(\mathcal{P}(\triangleright)) \neq \mathcal{H}(\mathcal{P}(\triangleright))$, it indicates tampering, and $s$ can reject the request.
    \end{enumerate}
    For \textbf{Response Integrity},
    \begin{enumerate}
        \item After processing the request, $s$ encrypts the raw response $\triangleleft$ with a symmetric key $\kappa$ to produce $\bar{\triangleleft} = \textit{Enc}(\triangleleft, \kappa)$.
        \item $s$ then computes the cryptographic hash $H(\bar{\triangleleft})$ of this encrypted response.
        \item $H(\bar{\triangleleft})$ is committed on $\mathfrak{B}$ as part of $\mathcal{X}(\triangleleft) = \anglebrackets{\mathcal{H}(\mathcal{X}(\mathcal{P}(\triangleright))), H(\bar{\triangleleft}), \eta}$. Again, $\mathfrak{B}$'s immutability guarantees the integrity of this committed hash.
        \item $u$ retrieves $\bar{\triangleleft}$ from the off-chain storage facility $\mathcal{D}(\bar{\triangleleft})$ and independently computes its hash, $H'(\bar{\triangleleft})$.
        \item $u$ verifies the integrity of the retrieved encrypted response by comparing $H'(\bar{\triangleleft})$ with $H(\bar{\triangleleft})$ from $\mathfrak{B}$. A mismatch indicates tampering with the encrypted response.
    \end{enumerate}
    Therefore, any unauthorized alteration of request or response payloads, whether in transit or storage, will be detected through cryptographic hash mismatches verified against the immutable ledger, thus ensuring communication integrity.
\end{proof}

\begin{property}[Authenticity and Non-Repudiation]
    Senders of requests and responses can be verifiably identified, and they cannot falsely deny having originated these messages.
\end{property}

\begin{proof}
    We first prove \textbf{Authenticity}.
    \begin{enumerate}
        \item Each PA $u$ and SA $s$ is assigned a unique DID, $\textit{DID}(u)$ and $\textit{DID}(s)$ respectively.
        \item These DIDs are registered and managed on the VAR, which is implemented as a smart contract on $\mathfrak{B}$.
        \item The cryptographic nature of DIDs and the immutability of the VDR on $\mathfrak{B}$ ensure that each agent's identity is verifiably linked to its DID, and that DIDs cannot be forged or impersonated without compromising the underlying cryptographic keys.
    \end{enumerate}
    We prove the non-repudiation via request non-repudiation and response non-repudiation, respectively.
    For \textbf{Request Non-Repudiation},
    \begin{enumerate}
        \item The request transaction $\mathcal{X}(\mathcal{P}(\triangleright))$ explicitly includes $\textit{DID}(u)$ (the sender's identifier) and $\textit{DID}(s)$ (the intended recipient's identifier), along with the cryptographic hash of the request payload $\mathcal{H}(\mathcal{P}(\triangleright))$.
        \item $\mathcal{X}(\mathcal{P}(\triangleright))$ is broadcasted to and recorded on $\mathfrak{B}$. Given that transactions on a blockchain are cryptographically signed by the originating entity, $u$ cannot subsequently deny having initiated the request. The on-chain record serves as irrefutable proof.
    \end{enumerate}
    For \textbf{Response Non-Repudiation},
    \begin{enumerate}
        \item The response transaction $\mathcal{X}(\triangleleft)$ includes $\mathcal{H}(\mathcal{X}(\mathcal{P}(\triangleright)))$ (linking it to the specific request) and $H(\bar{\triangleleft})$ (the hash of the encrypted response).
        \item $\mathcal{X}(\triangleleft)$ is broadcasted by $s$ to $\mathfrak{B}$ and recorded immutably.
        \item By committing $\mathcal{X}(\triangleleft)$ to $\mathfrak{B}$, $s$ publicly attests to having processed the request and generated the corresponding (encrypted) response. $s$ cannot later deny having sent this response, as its action is verifiably recorded on $\mathfrak{B}$.
    \end{enumerate}
    Thus, the combination of cryptographically verifiable DIDs and immutable on-chain transaction records provides strong guarantees for authenticity and non-repudiation for both request and response commitments.
\end{proof}

\begin{property}[Response Confidentiality]
    The raw response $\triangleleft$ is only revealed to the authorized PA $u$ after specific conditions $\eta$ are verifiably met.
\end{property}

\begin{proof}
We assume the correctness and robustness of cryptographic algorithms in our proof.
    \begin{itemize}
    \item \textbf{Raw Response Encryption:} $s$ encrypts the raw response $\triangleleft$ using a randomly generated symmetric encryption key $\kappa$, resulting in $\bar{\triangleleft} = \textit{Enc}(\triangleleft, \kappa)$. $\bar{\triangleleft}$ is then stored off-chain in $\mathcal{D}(\bar{\triangleleft})$. Without knowledge of $\kappa$, $\bar{\triangleleft}$ is computationally infeasible to decrypt, preserving $\triangleleft$ confidentiality.
    \item \textbf{Conditional Key Release:} $\kappa$ is not directly transmitted. Instead, $s$ encrypts $\kappa$ using $u$'s public key $\textit{PK}(u)$, yielding $\bar{\kappa} = \textit{Enc}(\kappa, \textit{PK}(u))$. $\bar{\kappa}$ is released by $s$ to $u$ only after the conditions $\eta$ (specified in $\mathcal{X}(\triangleleft)$ on $\mathfrak{B}$) are verifiably fulfilled on $\mathfrak{B}$. The on-chain verification mechanism for $\eta$ (e.g., smart contract logic for payment) ensures this conditional release.
    \item \textbf{Asymmetric Decryption:} Only $u$ possesses the corresponding private key $\textit{SK}(u)$ to $\textit{PK}(u)$. Therefore, only $u$ can decrypt $\bar{\kappa}$ to retrieve $\kappa$ ($\kappa = \textit{Dec}(\bar{\kappa}, \textit{SK}(u))$). Any unauthorized entity attempting to intercept $\bar{\kappa}$ would be unable to decrypt it without $\textit{SK}(u)$.
    \item \textbf{Raw Response Decryption:} Once PA $u$ obtains $\kappa$, it can then decrypt the off-chain stored $\bar{\triangleleft}$ to reveal the raw response $\triangleleft$ ($\triangleleft = \textit{Dec}(\bar{\triangleleft}, \kappa)$).
\end{itemize}
The multi-layered encryption scheme, combined with the on-chain enforcement of conditions for key release, ensures that the confidentiality of the raw response $\triangleleft$ is maintained until the precise moment and to the precise entity ($u$) for which access is authorized and conditions are fulfilled.
\end{proof}

\begin{property}[Verifiable Condition Fulfillment]
    The conditions $\eta$ for releasing the decryption key $\kappa$ are verifiably fulfilled on $\mathfrak{B}$.
\end{property}

\begin{proof}
    We formulate the proof outline as follows.
    \begin{itemize}
    \item \textbf{On-Chain Declaration:} The conditions $\eta$ are an integral part of the response commitment transaction $\mathcal{X}(\triangleleft)$ that SA $s$ broadcasts to $\mathfrak{B}$, meaning that $\eta$ is publicly declared and immutably recorded on the ledger before any key release occurs. This transparency allows any interested party to inspect the terms of engagement.
    \item \textbf{Smart Contract Enforcement:} If $\eta$ involves a verifiable action (e.g., payment, proof of data access, computational proof), this action is executed and verified by a smart contract deployed on $\mathfrak{B}$.
    \item \textbf{Irrefutable Proof of Fulfillment:} The successful execution of the action specified by $\eta$ results in a state change on $\mathfrak{B}$ (e.g., a token balance update, an event emission by a smart contract). This on-chain state change serves as an irrefutable, timestamped, and publicly auditable proof that the conditions $\eta$ have been met.
    \item \textbf{On-Chain Verification:} SA $s$ is designed to release the encrypted symmetric key $\bar{\kappa}$ to PA $u$ only after it has verified the on-chain fulfillment of $\eta$. This verification process involves querying the state of $\mathfrak{B}$ to confirm the necessary transactions or state changes have occurred, ensuring that SA $s$ is compensated or its conditions are met before revealing the sensitive data.
\end{itemize}
Therefore, the DMAS guarantees verifiable condition fulfillment by leveraging the transparency and immutability of the distributed ledger, making all conditional agreements and their resolution publicly auditable and trustless.

\end{proof}

\section{Performance Analysis}
\label{sec:performance_analysis}

We evaluate the scalability and efficiency of the DMAS under various operational conditions.

\subsection{Experimental Setting}
\label{sec:experimental_setting}

\begin{itemize}
    \item \textbf{Blockchain Network ($\mathfrak{B}$):} A local instance of an Ethereum testnet was deployed using Hardhat Network for controlled and repeatable experiments. The block time was configured to 2 seconds to accurately simulate a Layer-2 network environment, such as Base, known for its faster block finality compared to Ethereum's mainnet.
    \item \textbf{DMAS:} Both PAs and SAs were implemented as Python programs leveraging Autogen (v0.6.1) for robust multi-agent conversation and orchestration capabilities. Notably, all agents (PAs and SAs) utilized GPT-4o as their underlying LLM for reasoning, natural language understanding, and response generation. The experimental setup involved deploying these agents within Docker containers for isolation and streamlined resource management. The number of PAs was varied from $1$ to $32$ concurrent instances to simulate increasing load. A diverse pool of $32$ distinct SAs was pre-registered in the VAR. These SAs are categorized into $4$ groups, with each group containing $7$ unique terminal agents, each discoverable and routable by a specific routing agent. To initiate service discovery, each PA makes $8$ distinct requests, initially engaging all $4$ routing agents.
    \item \textbf{Centralized MAS:} For the comparative experiment, we developed a Centralized MAS (CMAS) by adapting the DMAS implementation. We orchestrate SAs into two layers. The first layer comprises $4$ routing agents, while the second layer consists of the other $28$ agents. Unlike the DMAS where PAs interact with SAs via the blockchain, each PA is pre-configured to connect directly to all $4$ routing agents in the first layer. The PAs then iteratively call their underlying LLMs (GPT-4o, as specified in the DMAS setup) to process requests and manage the flow of interaction through these two layers.
    \item \textbf{Hardware Environment:} All experiments were executed on a single powerful physical machine, equipped with an Intel Core Ultra 9 285K processor, 64GB of RAM, and an RTX 5090 GPU. For the DMAS setting, this machine hosted the $32$ Docker containers, with each container isolating a PA and an SA, allowing for a controlled assessment of resource utilization and performance under load. The CMAS is encapsulated in a single container for the comparative experiment.
\end{itemize}

\subsection{Experiment Design}

We first conducted a scalability experiment by gradually increasing the number of concurrent PAs from 1 to 32, allowing us to observe and analyze how the percentage distribution of on-chain versus off-chain time costs shifted under varying loads. Our goal was to understand how the DMAS's performance characteristics evolved as more agents became active.

To isolate and quantify the performance overhead specifically attributable to the decentralized mechanisms of the DMAS, we further conducted a comparative experiment with the CMAS described in Section~\ref{sec:experimental_setting}. The CMAS bypasses all blockchain interactions, including DID resolution, verifiable commitments, and conditional key release, relying instead on pre-configured direct connections and a two-layered agent orchestration model.

\subsection{Scalability Analysis}

The total end-to-end latency for a full DMAS interaction cycle (from PA request initiation to decrypted response) can be broadly decomposed into on-chain and off-chain time cost.

\begin{itemize}
    \item \textbf{On-chain Time Cost} primarily includes the time spent waiting for blockchain transaction confirmations for the necessary verifiable steps, such as request commitment ($\mathcal{X}(\mathcal{P}(\triangleright))$), response commitment ($\mathcal{X}(\triangleleft)$), and condition fulfillment (e.g., payment verification). These operations are inherently bound by the block time, consensus mechanism overhead, and network propagation delays.
    \item \textbf{Off-chain Time Cost} includes all other delays. This includes the actual computation performed by SAs, the network latency for direct peer-to-peer payload transfers between PAs and SAs, the time taken for service discovery iterations (including any delegation hops), the latency for calling external LLM services (specifically OpenAI APIs for GPT-4o in our setup), and the subsequent data processing by agents.
\end{itemize}

\begin{figure}[htbp]
\centerline{\includegraphics[width=0.9\columnwidth]{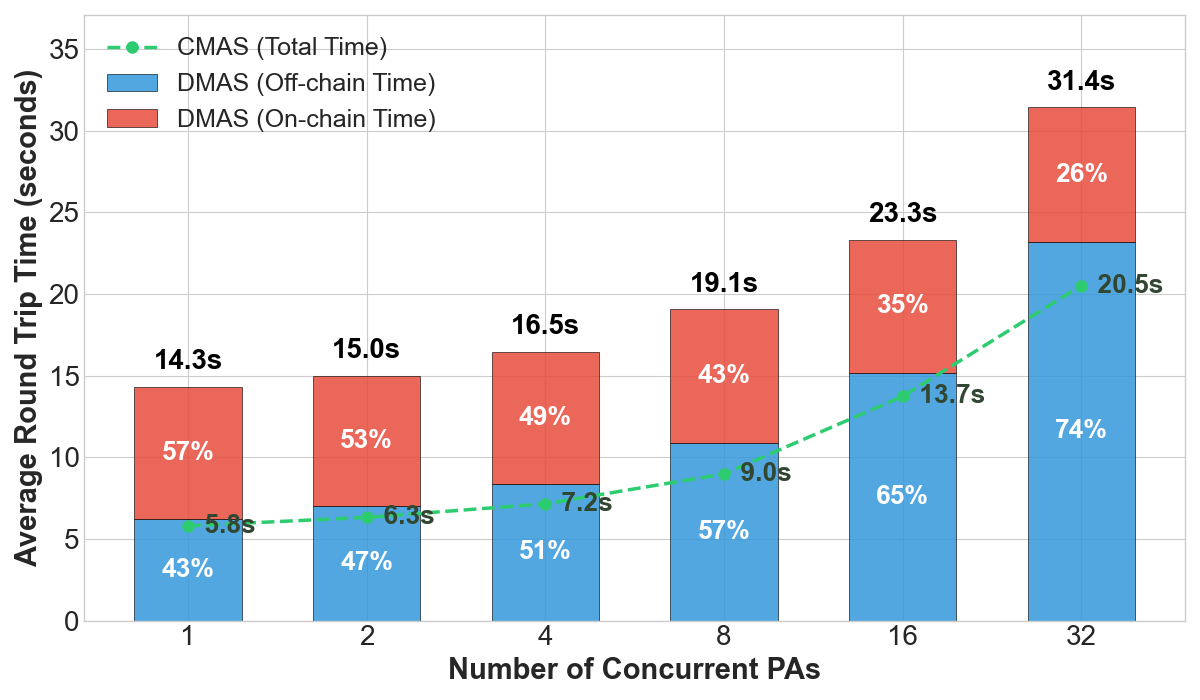}}
\caption{Performance experiment results.}
\label{fig:performance}
\end{figure}

Our experimental results, as illustrated in Figure~\ref{fig:performance}, demonstrate that on-chain time cost is a significant, but decreasing, percentage of the total time cost as off-chain operations scale with the number of concurrent requests.

The observed trend highlights that the DMAS effectively pushes the performance bottleneck away from the blockchain for a wide range of workloads. The hybrid design allows the DMAS to leverage the high throughput capabilities of traditional off-chain distributed computing for complex agent interactions, while relying on the blockchain only for critical verifiable attestations.

\subsection{Efficiency Analysis}

The line chart in Figure~\ref{fig:performance} illustrates the end-to-end latency for the CMAS across varying numbers of concurrent requests.

A critical finding emerges when we compare the total latency of the CMAS with the DMAS's total latency after subtracting its on-chain time cost. As depicted by the nearly identical curves compared to the DMAS off-chain time cost in Figure~\ref{fig:performance}, our experiments lead to a conclusion: excluding the on-chain time cost, the DMAS shares almost the same efficiency compared to the CMAS.

This direct comparison unequivocally demonstrates that the DMAS mimics the computational and communication efficiency of a centralized system for its off-chain operations. The architectural choice to offload heavy computation from the blockchain to a distributed off-chain environment is thus validated as highly effective. The observed performance difference in total latency between the DMAS and CMAS is, almost entirely, a direct consequence of the blockchain interactions required for ensuring the security properties.

\section{Conclusion}

We have presented the Decentralized Multi-Agent System (DMAS), a novel architectural paradigm designed to overcome the inherent limitations of centralized multi-agent systems, such as single points of failure, censorship, and trust deficits, by integrating a decentralized agent runtime and a trust-aware communication protocol. Our security analysis formally validates the system's ability to ensure message integrity, agent authenticity, non-repudiation, and confidential data exchange. Furthermore, the performance analysis demonstrates the DMAS's practical scalability and efficiency, underscoring its potential for fostering a trustworthy and scalable Internet of Agents.

\bibliographystyle{IEEEtran}
\bibliography{IEEEabrv,references}

\end{document}